# Correlations between stacked structures and weak itinerant magnetic properties of La$_{2-x}$Y$_x$Ni$_7$ compounds

Valérie Paul-Boncour[1*], Jean-Claude Crivello[1], Nicolas Madern[1#], Junxian Zhang[1], Léopold V.B. Diop[2], Véronique Charbonnier[1,3], Judith Monnier[1], Michel Latroche[1]

[1]Univ. Paris-Est Créteil, CNRS, ICMPE, UMR7182, F-94320 Thiais, France

[2]Université de Lorraine, CNRS, IJL, F-54000 Nancy, France

[3]Energy Process Research Institute, National Institute of Advanced Industrial Science and Technology (AIST), Tsukuba West, 16-1 Onogawa, Tsukuba, Ibaraki 305-8569, Japan.

**Abstract**

Hexagonal La$_2$Ni$_7$ and rhombohedral Y$_2$Ni$_7$ are weak itinerant antiferromagnet (wAFM) and ferromagnet (wFM), respectively. To follow the evolution between these two compounds, the crystal structure and magnetic properties of $A_2B_7$ intermetallic compounds ($A$ = La, Y, $B$ = Ni) have been investigated combining X-ray powder diffraction and magnetic measurements. The La$_{2-x}$Y$_x$Ni$_7$ intermetallic compounds with $0 \leq x \leq 1$ crystallize in the hexagonal Ce$_2$Ni$_7$-type structure with Y preferentially located in the [$A_2B_4$] units. The compounds with larger Y content (1.2 $\leq x <$ 2) crystallize in both hexagonal and rhombohedral (Gd$_2$Co$_7$-type) structures with a substitution of Y for La in both [$A_2B_4$] and [$AB_5$] units. Y$_2$Ni$_7$ crystallizes in the rhombohedral structure only. The average cell volume decreases linearly *versus* Y content, whereas the $c/a$ ratio presents a minimum at $x = 1$ due to geometric constrains. The magnetic properties are strongly dependent on the structure type and the Y content. La$_2$Ni$_7$ displays a complex metamagnetic behavior with split AFM peaks. Compounds with $x = 0.25$ and 0.5 display a wAFM ground state and two metamagnetic transitions, the first one towards an intermediate wAFM state and the second one towards a FM state. $T_N$ and the second critical field $\mu_0H_{c2}$ increase with the Y content, indicating a stabilization of the AFM state. LaYNi$_7$, which is as the boundary between the two structure types, presents a very wFM state at low field and an AFM state as the applied field increases. All the compounds with $x > 1$, and which contains a rhombohedral phase are wFM with $T_C$ = 53(2) K. In addition to the experimental studies, first principles calculations using spin polarization have been performed to interpret the evolution of structural phase stability for $0 \leq x \leq 2$.

[*]Corresponding author : paulbon@icmpe.cnrs.fr, Phone number:+33 1 49 78 12 07

[#] Current affiliation : CEA SACLAY, 91191 Gif-sur-Yvette Cedex, France

## 1. Introduction

$A_2Ni_7$ compounds ($A$ = Rare Earth) have raised interest for their fundamental physical properties [1-3] as well as their applications as hydrogen storage materials [4-6] and negative electrodes in Ni-$M$H batteries [7-11]. $A_2Ni_7$ intermetallic compounds crystallize in two polymorphic crystal structures which are either hexagonal ($Ce_2Ni_7$-type, $P6_3/mmc$) or rhombohedral ($Gd_2Co_7$-type, $R$-$3m$) depending on the stacking of the [$A_2B_4$] and [$AB_5$] units along the $c$ axis ($B$ = Ni) [1]. The hexagonal ($2H$) structure is favored for intermetallic compounds with large $A$ elements (La, Ce) and the rhombohedral ($3R$) for smaller $A$ ones (Ho, Er, Y) [1, 3]. A mixture of both polymorphs is obtained for the compounds which contain intermediate $A$ size elements. However, the annealing temperature can also be used to tune the weight percentage of each phase [1, 12].

Their magnetic properties have been investigated [13], and particular interest has been raised on the peculiar itinerant magnetic properties of $La_2Ni_7$ [12, 14-17] and $Y_2Ni_7$ [18-23] which both contain non-magnetic $A$ elements. Weak itinerant magnetism is related to the electronic structure and the small magnetic moment arises from the exchange splitting of the bands near the Fermi level. Weak itinerant magnets are close to the boundary between magnetic and non-magnetic state, and a small structural variation can induce large changes in both electronic and magnetic properties. Such magnetic instabilities have been observed and studied in other weak ferromagnets (wFM) as $Ni_3Al$ [24-26], $Fe_2N$ [27, 28], $AsNCr_3$ [29], $CrAlGe$ [30] or weak antiferromagnets (wAFM) like $TiAu$ [31] and $UN$ [32]. The weak itinerant magnetism is often associated to large spin fluctuations. For example, although $Ni_3Al$ and $Ni_3Ga$ have close electronic structures, they differ by their magnetic properties: $Ni_3Al$ is a weak itinerant ferromagnet, whereas $Ni_3Ga$ is a paramagnet [26, 33]. This difference has been attributed to the larger spin fluctuations in $Ni_3Ga$ [26].

Earlier reported magnetic investigations on hexagonal $La_2Ni_7$ have shown that the ground state is a weak itinerant antiferromagnet with a Néel temperature $T_N$ = 50 K at low field [12, 34]. Multiple field-induced metamagnetic transitions from wAFM state towards a weak itinerant ferromagnetic state have been observed. The transition fields-temperature phase diagram reveals the existence of different magnetic transitions below 60 K [14-17]. Parker *et al* [12] were able to obtain a sample containing both hexagonal and rhombohedral structures after annealing at 873 K, and found a wFM behavior with a Curie temperature $T_C$ = 70 K associated to rhombohedral $La_2Ni_7$. In both hexagonal and rhombohedral phases, the saturation magnetization $\mu_S$ remains weak (0.11 and 0.086 $\mu_B$/Ni respectively), whereas the effective Ni moment determined by a Curie Weiss law in the paramagnetic range varies between 0.9 and 1.04 $\mu_B$/Ni. In a recent theoretical work, an hexagonal AFM structure has been proposed, described by two FM unit blocks of opposite Ni spin sign separated by a non-magnetic layer at $z = 0$ and ½ [35]. This AFM structure and the corresponding FM structure are more stable with Ni moments parallel to the $c$ axis. The Ni moment magnitude depends on the neighboring atoms, it is minimum in the [$A_2B_4$] unit and increase to a maximum at the interface between two [$AB_5$] units.

Rhombohedral $Y_2Ni_7$ has been described as a wFM with $T_C$ = 53 to 60 K and $\mu_S$ = 0.06 to 0.08 $\mu_B$/Ni, depending on the Ni content ($Y_2Ni_{6.7}$ to $Y_2Ni_7$) [19-22, 36]. A theoretical study of its electronic density has been undertaken to explain the origin of the itinerant ferromagnetic properties of $Y_2Ni_7$ with a very weak Ni moment and a corresponding relatively elevated transition temperature [23]. A sharp and narrow peak is observed in the density of state (DOS) at the Fermi level ($E_F$) leading to a high density of state and stabilizing the spin polarized configuration with a ferromagnetic state.

The influence of partial substitution of Gd for Y [37], Al, Co and Cu for Ni [15, 38, 39] as well as hydrogen insertion [34, 36] has revealed a large sensitivity of the magnetic properties for both $La_2Ni_7$ and $Y_2Ni_7$ to these chemical substitutions or insertion. However, it appears to have no studies on the properties of intermediate compounds between $La_2Ni_7$ and $Y_2Ni_7$. Due to the difference of crystal structure (hexagonal/rhombohedral) and itinerant magnetic properties (antiferromagnetic / ferromagnetic) of the binary compounds, it is worth to understand how the change between these different states is occurring: is it continuous or discontinuous, is there any critical Y concentration at which the change of crystal structure occurs? One important question is the correlation between the crystal structure and the magnetic order. In previous DFT calculation study [35], we have found an AFM structure which is compatible with the hexagonal structure, but not with the rhombohedral one. In addition, for hexagonal $La_2Ni_7$, both AFM and FM structures have comparable magnetic energy at 0 K, which raises the question of stability range of the AFM state and how it is modified by structural changes.

To answer these questions, we have investigated the structural and magnetic properties of the $La_{2-x}Y_xNi_7$ system ($0 \leq x \leq 2$) to follow the evolution from the hexagonal $La_2Ni_7$ antiferromagnet towards the rhombohedral $Y_2Ni_7$ ferromagnet. Complementary to the experimental study, band structure calculations have been undertaken to interpret the evolution of these properties. This will contribute to determine the key geometric and electronic parameters which drive the change of the physical properties of these compounds.

## 2. Experimental and calculation conditions

Polycrystalline samples with composition $La_{2-x}Y_xNi_7$ ($x$ = 0, 0.25, 0.5, 1, 1.2, 1.4 and 2) were prepared by induction melting under a purified argon atmosphere and annealed 7 days at 1223-1273 K as described in previous work [5]. The samples were annealed at high temperature to get compounds with a single structure type when possible.

The samples were characterized by X-ray diffraction (XRD) using a D8 diffractometer from Bruker with Cu $K_\alpha$ radiation. The XRD patterns were refined using the Fullprof code [40]. Their chemical compositions were analyzed by electron probe microanalysis (EPMA) using a SX100 from CAMECA.

The magnetic measurements were performed on small quantity of sample (bulk piece or powder fixed by a resin) with a Physical Properties Measurement System (Quantum Design PPMS-9) operating from 2 to 300 K, with a maximum applied field of 9 T. The magnetization measurements of $La_{1.75}Y_{0.25}Ni_7$ and $La_{1.5}Y_{0.5}Ni_7$ were extended up to 14 T using a commercial vibrating-sample magnetometer (Quantum Design PPMS-14).

The electronic structures were calculated for the ordered compounds $La_2Ni_7$ and $Y_2Ni_7$ in both hexagonal ($H$) and rhombohedral ($R$) symmetries. Since no distinct total energy differences were observed in the non-spin polarized calculations between binaries with $R$ or $H$ structure, the $La_{2-x}Y_xNi_7$ compounds with $x$ = 0, 0.5, 1, 1.5 and 2 have been only considered in the rhombohedral cell (primitive rhombohedral, not hexagonal for CPU-time saving). In this 18-atoms cell, all ordered kinds of $A$-site configurations have been considered ($6c_1$ and $6c_2$ with half or full substitution of La by Y). This $A$-ordering approximation differs from an ideal solid solution representation where a random distribution is preferable, but is sufficient since our preliminary tests on pseudo-disordered supercells presented similar energetic results that the ordered case [41].

In the frame of the DFT, the pseudo-potential approach using the VASP package [42, 43] was considered by applying the generalized gradient approximation with the PBE functional [44] using a 600 eV cut-off energy and a high $k$-mesh density (grid of 9x9x9 for the rhombohedral and 21x21x7 for the hexagonal structure). Preserving the original crystal symmetry, each structure has been fully relaxed with the electronic collinear and non-collinear spin-polarization.

## 3. Results and discussion

### 3.1. Structural properties

The $A_2Ni_7$ alloys can adopt either the rhombohedral $Gd_2Co_7$-type or the hexagonal $Ce_2Ni_7$-type structure, as shown in Figure 1. Both structures are related to the stacking along the *c* axis of $[A_2B_4]$ and $[AB_5]$ units according to the rule $[A_2B_4]+n[AB_5]$, where *n* is an integer. Note that the $[A_2B_4]$ subunit will be renamed as $[AB_2]$ for simplicity in the following. The rhombohedral cell (hexagonal description) contains three stacking of two $[AB_5]$ and one $[AB_2]$ units, whereas the hexagonal cell contains only two of them, with a mirror inversion of the $[AB_2]$ units located in the top $[AB_2']$. In each cell, the *A* atoms belong to two different Wyckoff sites (Table 1): the *A* atom belonging to the $[AB_2]$ unit is surrounded by 16 atoms (CN16, with CN for Coordination Number), whereas the one belonging to the $[AB_5]$ unit is surrounded by 20 atoms (CN20).

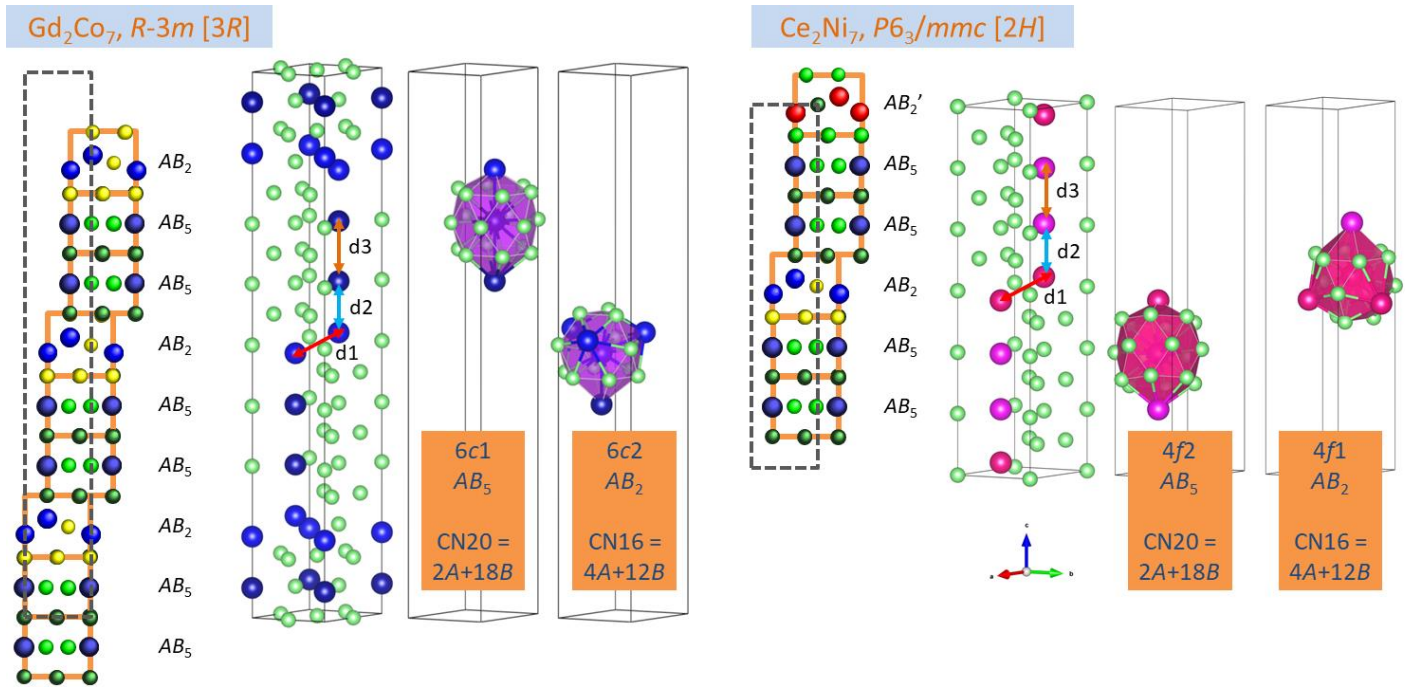

**Figure 1**. Illustration of the rhombohedral $Gd_2Co_7$-type and hexagonal $Ce_2Ni_7$-type structures with the coordination spheres of each *A* Wyckoff sites. The three different *A*–*A* interatomic distances ($d_1$, $d_2$, $d_3$) are indicated by arrows.

The XRD patterns were refined with a single hexagonal structure ($Ce_2Ni_7$-type structure) for $x \leq 1$, a mixture of hexagonal and rhombohedral structures for $1 < x < 2$ and a single rhombohedral structure ($Gd_2Co_7$-type structures) for $x = 2$ in agreement with [22, 36]. The refined patterns of samples with $x = 0, 0.25, 1, 1.2$ are presented in supplementary material (Figure S1) and the factors of agreement are added in the legend. The crystal structure analysis of rhombohedral $Y_2Ni_7$ is presented in details in ref. [5]. All the hexagonal compounds are well crystallized, without anisotropic line broadening.

Note that the XRD pattern of $La_{1.75}Y_{0.25}Ni_7$ displays small additional lines due to 1.4 wt% of cubic $LaNi_2$, whereas that of $La_{0.8}Y_{1.2}Ni_7$ the few additional lines are refined with 1.7 wt% of $Y_2O_3$.

In addition, difference between the calculated and experimental patterns for $x = 1.2$ and 1.4 is observed in the $2\theta$ range between 25 and 35° (inset of Figure S1b). This difference can be attributed to the existence of microstructural defects. These defects have been interpreted as stacking faults due to some random local variation of $n$ in the stacking of the $[A_2B_4]$ +$n$ $[AB_5]$ units along the $c$ axis. These defects have been clearly observed by HAADF STEM images in $La_{0.65}Nd_{0.15}Mg_{0.20}Ni_{3.5}$ and $La_{0.75-0.80}Mg_{0.30-0.38}Ni_{3.67}$ alloys [45, 46]. Such defects cannot be considered by classical diffraction analysis as they are not periodic. Indeed, the diffraction patterns show some broadening located between 30 and 35° that are directly related to these stacking faults and that cannot be refined using Fullprof refinement. A more detailed analysis of these stacking faults is under progress with the FAULTS code [47] and will be detailed in a further work, but a first simulation indicates a maximum of 9% of stacking faults in $Y_{1.2}La_{0.8}Ni_7$. It is important to note that such defects have small incidence on the diffraction patterns as most of the X-ray diffraction peaks are not affected by these stacking faults. Therefore, the full pattern refinement performed with Fullprof fully allows to determine the atomic positions and phase contents.

The cell parameters as well as the chemical compositions determined by EPMA for the $La_{2-x}Y_xNi_7$ compounds are summarized in Table 2. The evolution of $a$ and $c$ cell parameters, cell volume $V$ and $c/a$ ratio *versus* Y content is reported in Figure 2. The weight percentage of each phase is shown in Figure 3a. For $x = 1.2$ and 1.4, the $a_R$ and $a_H$ cell parameters are similar, whereas the rhombohedral $c_R$ parameter is 3/2 times the hexagonal $c_H$ one. The $a$ ($a_H$ and $a_R$) and $c$ ($c_H$ and 2/3 $c_R$) cell parameters decrease continuously *versus* Y content following a second order polynomial law with opposite curvatures (the polynomial coefficients are reported in Figure 2). The $c/a$ ratio (hexagonal description) passes through a minimum near $x = 1$, then increases again. For $x < 1$, the $c/a$ ratio follows a power law with a square exponent ($y(c/a) = 0.27(x-1)^2$). On the other hand, a linear decrease of the reduced cell volume $V/Z$ with $d(V/Z)/dx = 2.97$ Å$^3$ f.u.$^{-1}$ is observed independently of the structure type, in agreement with the Vegard's law expected for a solid solution (Figure 2).

**Table 1.** Tabulated atomic positions for $A_2Ni_7$ phases in hexagonal ($P6_3/mmc$) and rhombohedral ($R-3m$) symmetries.

| Hexagonal | $Ce_2Ni_7$ | Type | | | | | |
|---|---|---|---|---|---|---|---|
| atom | Wyckoff | $x$ | $y$ | $z$ | Unit | Neighbors | CN |
| A1 | $4f_1$ | 1/3 | 2/3 | 0.5 | $[AB_2]$ | 4 $A$+12 Ni | 16 |
| A2 | $4f_2$ | 1/3 | 2/3 | 0.667 | $[AB_5]$ | 2 $A$+18 Ni | 20 |
| Ni1 | $12k$ | 0.834 | $2x$ | 0.085 | $[AB_2]/[AB_5]$ | 5 $A$+7 Ni | 12 |
| Ni2 | $6h$ | 0.167 | $2x$ | 1/4 | $[AB_5]$ | 4 $A$+8 Ni | 12 |
| Ni3 | $4f_3$ | 1/3 | 2/3 | 0.167 | $[AB_5]$ | 3 $A$+9 Ni | 12 |
| Ni4 | $4e$ | 0 | 0 | 0.167 | $[AB_5]$ | 3 $A$+9 Ni | 12 |
| Ni5 | $2a$ | 0 | 0 | 0 | $[AB_2]$ | 6 $A$+6 Ni | 12 |
| Rhombohedral | $Gd_2Co_7$ | Type | | | | | |
| atom | Wyckoff | $x$ | $y$ | $z$ | Unit | | |
| A1 | $6c_1$ | 0 | 0 | 0.056 | $[AB_5]$ | 2 $A$+18 Ni | 20 |
| A2 | $6c_2$ | 0 | 0 | 0.167 | $[AB_2]$ | 4 $A$+12 Ni | 16 |
| Ni1 | $18h$ | 0.5 | $-x$ | 0.111 | $[AB_2]/[AB_5]$ | 5 $A$+7 Ni | 12 |
| Ni2 | $9e$ | ½ | 0 | 0 | $[AB_5]$ | 4 $A$+8 Ni | 12 |
| Ni3 | $6c_3$ | 0 | 0 | 0.278 | $[AB_5]$ | 3 $A$+9 Ni | 12 |
| Ni4 | $6c_4$ | 0 | 0 | 0.389 | $[AB_5]$ | 3 $A$+9 Ni | 12 |
| Ni5 | $3b$ | 0 | 0 | ½ | $[AB_2]$ | 6 $A$+6 Ni | 12 |

**Table 2.** Chemical composition obtained from EPMA and cell parameters obtained from Rietveld refinement of the $La_{2-x}Y_xNi_7$ compounds.

| | | Hexagonal-2$H$ | | | | Rhombohedral-3$R$ | | | |
|---|---|---|---|---|---|---|---|---|---|
| $x$ | EPMA | $a$ (Å) | $c$ (Å) | $V$ (Å$^3$) | wt. % | $a$ (Å) | $c$ (Å) | $V$ (Å$^3$) | wt. % |
| 0 | $La_2Ni_{6.86}$ | 5.0627(1) | 24.714(7) | 548.57(2) | 100 | | | | |
| 0.25 | $Y_{0.23}La_{1.77}Ni_{6.80}$ | 5.0540(1) | 24.630(1) | 544.81(1) | 100 | | | | |
| 0.5 | $Y_{0.45}La_{1.55}Ni_{6.94}$ | 5.0405(2) | 24.509(1) | 539.26(4) | 100 | | | | |
| 1 | $Y_{1.01}La_{0.99}Ni_{6.86}$ | 5.0162(3) | 24.359(2) | 530.81(6) | 100 | | | | |
| 1.2 | $Y_{1.18}La_{0.82}Ni_{7.01}$ | 5.0062(1) | 24.324(5) | 527.90(2) | 33 | 5.0067(5) | 36.483(4) | 792.0(1) | 67 |
| 1.4 | $Y_{1.39}La_{0.61}Ni_{6.91}$ | 4.9937(4) | 24.273(2) | 524.20(8) | 27 | 4.9931(2) | 36.412(1) | 786.18(4) | 73 |
| 2 | $Y_2Ni_{6.92}$ | | | | | 4.9380(2) | 36.189(2) | 764.22(4) | 100 |

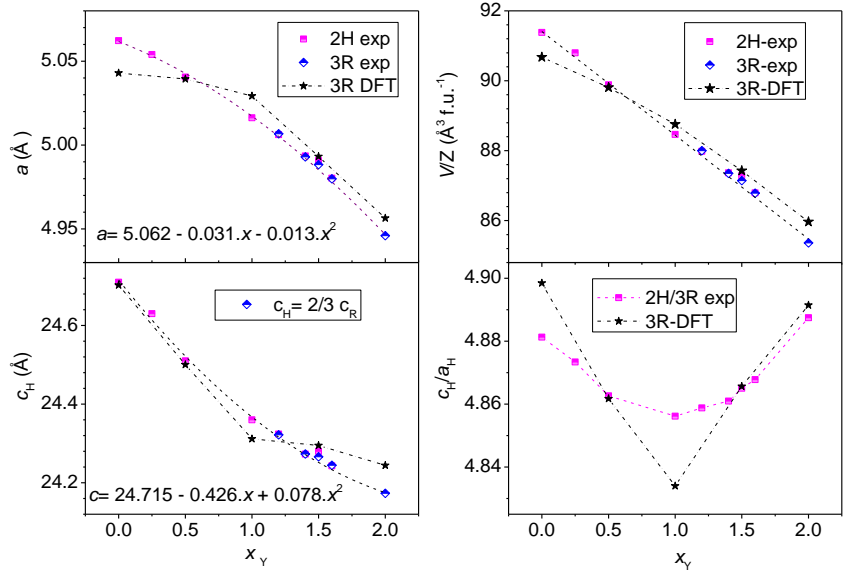

**Figure 2.**: Evolution of the experimental and DFT calculated $a$ and $c$ cell parameters, reduced cell volume $V/Z$ and $c/a$ ratio *versus* Y content in $La_{2-x}Y_xNi_7$.

The refined Y occupancy factors and atomic positions for each compound in both hexagonal and rhombohedral structures are reported in Table 3. The Y atoms are first substituted to La in the $4f_1$ site which belongs to the [$AB_2$] units with a coordination number CN16, where the compounds crystallize in hexagonal structure. For $x \geq 1$, an additional progressive filling of the $4f_2$ ($6c_1$) sites, belonging to the [$AB_5$] units (CN20), is observed in both hexagonal and rhombohedral structures (Figure 3b). The evolution of the $A$–$A$ interatomic distances presented in Figure 3c can be correlated to the cell parameter variations. First, we observe that the $A$–$A$ distances are shorter in the [$AB_2$] units ($d_1$) than in the [$AB_5$] units ($d_3$), whereas the distances between two $A$ atoms belonging to each unit type ($d_2$) are intermediate. The short $d_1$ distance decreases more sharply for $x > 1$, whereas the intermediate $d_2$ distance slightly increases. As the atoms belonging to the same [$AB_2$] units are almost in the same (*a*,*b*) plane whereas the *A* atoms belonging to two different units are aligned along the *c* axis (Figure 1), this interatomic distance variation reflects the inversion of the *c/a* ratio.

All the Ni atoms are surrounded by 12 atoms with different *A*:Ni ratios ranging from 6:6 for Ni5 atoms in the [$AB_2$] unit to 3:9 for Ni3 and Ni4 atoms in the [$AB_5$] unit. In La$_2$Ni$_7$, most of the Ni–Ni interatomic distances are between 2.45 and 2.62 Å, but significantly larger ones are obtained between Ni3 and Ni4 atoms with $d_{Ni-Ni}$ = 2.92 Å. Upon Y for La substitution, most of the Ni–Ni interatomic distances are decreasing except the $d_{Ni1-Ni3}$ and $d_{Ni1-Ni4}$, which are slightly increasing.

The DFT calculated *a*, *c*, *V/Z* parameters and *c/a* ratio have been added in Figure 2 (star symbols). Individual *a* and *c* parameters follow a variation comparable to the experimental ones with a small deviation according to the GGA approximation which overestimates Y-richer parameters. However, resultant values as the evolution of the relaxed cell volume and *c/a* ratio are both in rather good agreement with the experimental data, showing a continuous decrease of the cell volume and a minimum of the *c/a* ratio for *x* = 1. These calculations confirm that the substitution of Y for La in the [$AB_2$] unit favors a contraction of the *c* cell parameter whereas the filling of the [$AB_5$] unit favors a contraction of the *a* cell parameter. In Figure 2, it is observed that the calculated *c/a* ratio is significantly smaller than the experimental one for *x* = 1. Since the reported 0 K calculated values are taken from the ideal lower energy structure (Y only in [$AB_2$] units), this difference can be explained by the fact that experimentally there is a partial occupancy of Y substituted in both [$AB_2$] and [$AB_5$] sites (Table 3).

The *a*, *c*, *c/a* and *V/Z* cell parameters variations as well as the preferential Y site occupation among the two Wyckoff *A* sites calculated by DFT are in very good agreement with the experimental data. The anisotropic and inverse *c/a* variation *versus* Y content can be explained by the interfacial constraints between the two [$AB_5$] and [$AB_2$] units. For *x* < 1, the [$AB_5$] units contain only La atoms and the cell parameter variation upon Y substitution is mainly conducted by the [$AB_2$] contraction. As the interface between the [$AB_5$] and [$AB_2$] units is in the basal plane, the contraction of the [$AB_2$] units is limited in this direction to preserve the matching between the two types of units: the atom relaxation is therefore easier in the *c* direction for the [$AB_2$] units which are sandwiched between two [$AB_5$] units. When Y substitutes La in the [$AB_5$] unit, the cell parameters can relax in both *a* and *c* directions and the contraction in the basal plane becomes much easier to preserve the interface between the two units. Such anisotropic *c/a* variation has already been observed in La$_5$Ni$_{19}$ and Pr$_5$Ni$_{19}$ compounds upon Mg for La and Pr substitution [48, 49]. The Mg atoms replace the La (or Pr) atoms located in the [$AB_2$] units and the lattice contraction is much larger along the *c* axis than along the *a* axis. In Pr$_{3.75}$Mg$_{1.25}$Ni$_{19}$, the *c* cell parameter contraction (-1.5 %) is twice as high as for the parameter *a* (-0.6 %) [49].

The enthalpy of formation of the La$_{2-x}$Y$_x$Ni$_7$ compounds has been calculated considering the Y for La substitution in both possible interstitial sites ([$AB_5$] and [$AB_2$] units) using the rhombohedral description (Figure 4). Comparable results are expected for both polymorphic structures, as shown in [41]. In fact, no clear $\Delta H_{for}$ difference between both symmetries is observed for the binaries La$_2$Ni$_7$ and Y$_2$Ni$_7$. For *x* < 1, the position of Y in the 6$c_2$ sites ([$AB_2$] unit) leads to a more stable structure until this site is fully occupied by Y. Then for larger substitution rates of Y, a progressive filling of the 6$c_1$ site ([$AB_5$] unit) is observed. The Y atoms show therefore a preferential occupation in the less coordinated site (CN16) compared to the one with larger coordination number (CN20). This preferential occupation in CN16 site is attributed to the smaller atomic radius of Y atoms compared to the La ones. This CN16 site contains more *A* neighbors and less Ni neighbors, in agreement with the experimental observations as discussed above.

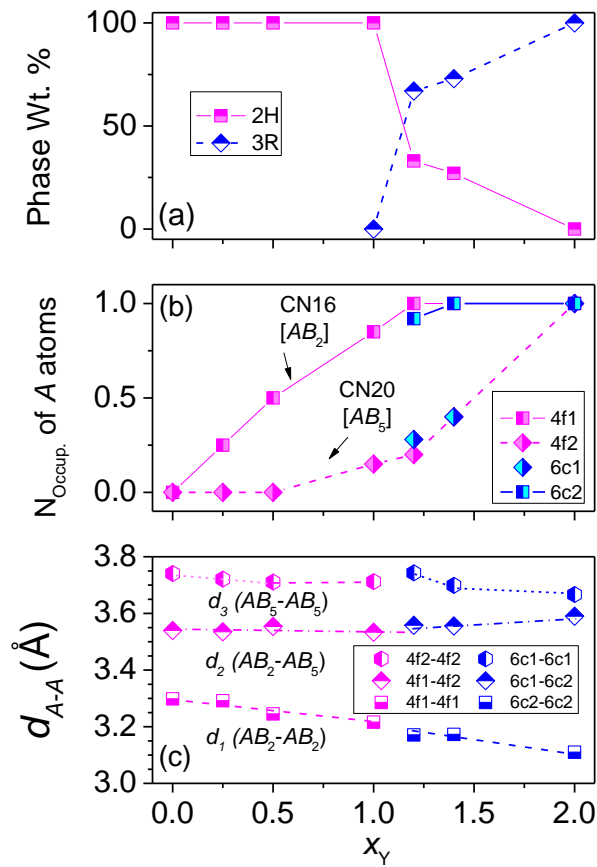

**Figure 3**. Evolution derived from Rietveld refinement of (a) the phase percentage in hexagonal and rhombohedral structures (b) $A$ site occupation by Y atoms and (c) $A$–$A$ interatomic distances ($d_1$, $d_2$ and $d_3$) as a function of the composition $La_{2-x}Y_xNi_7$.

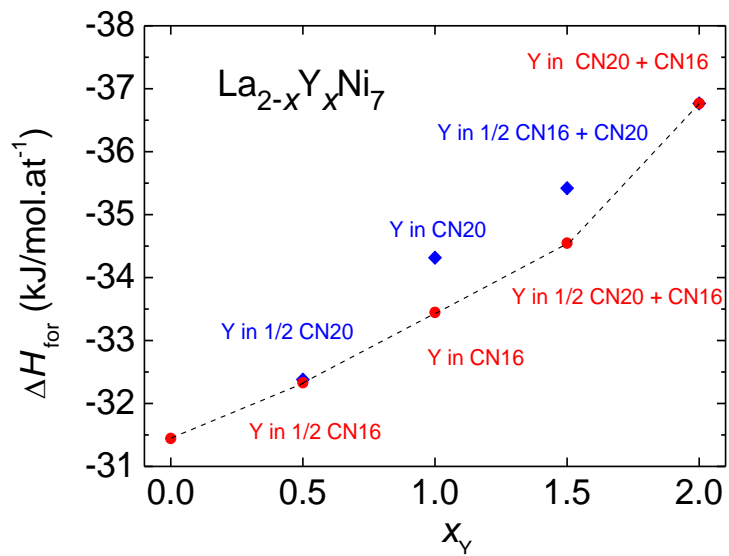

**Figure 4**. Enthalpies of formation for $La_{2-x}Y_xNi_7$ compounds calculated in the rhombohedral structure and for different $A$ sites ($6c_2$ in the [$AB_2$] units with CN16 and $6c_1$ in the [$AB_5$] units with CN20) occupancies.

**Table 3.** Occupancy factors and atomic positions derived from Rietveld refinement for $La_{2-x}Y_xNi_7$ compounds in hexagonal and rhombohedral structures.

| $x$ | 0 * | 0.25 | 0.5 | 1 | 1.2 | 1.4 | 2 |
|---|---|---|---|---|---|---|---|
| Structure | Hexagonal | | | | | | |
| $N_{occ}$ of Y in $4f_1$ [$AB_2$] | 0 | 0.25 | 0.5 | 0.85 | 1 | 1 | |
| $N_{occ}$ of Y in $4f_2$ [$AB_5$] | 0 | 0 | 0 | 0.15 | 0.2 | 0.4 | |
| $z$ (Y,La) $4f_1$ | 0.5300 (2) | 0.5309 (1) | 0.5293 (2) | 0.5286 (2) | 0.5107 (3) | 0.5280 (6) | |
| $z$ (Y,La) $4f_2$ | 0.6760 (3) | 0.6744 (1) | 0.6742 (2) | 0.6741 (2) | 0.6820 (2) | 0.6766 (5) | |
| $x$ Ni1 $12k$ | 0.169 (1) | 0.165 (1) | 0.166 (2) | 0.175 (1) | 0.176 (2) | 0.169 (4) | |
| $z$ Ni1 $12k$ | 0.0879(2) | 0.0873 (1) | 0.0867 (2) | 0.0847 (2) | 0.0850 (3) | 0.0841 (4) | |
| $x$ Ni2 $6h$ | 0.162(2) | 0.164 (1) | 0.176 (2) | 0.155 (3) | 0.156 (3) | 0.171 (6) | |
| $z$ Ni3 $4f_3$ | 0.1684(6) | 0.1686 (2) | 0.1671 (5) | 0.1682 (5) | 0.1792 (6) | 0.1700 (1) | |
| $z$ Ni4 $4e$ | 0.1713(6) | 0.1695(3) | 0.1707(6) | 0.1703(6) | 0.1592 (6) | 0.1731(1) | |
| Structure | Rhombohedral | | | | | | |
| $N_{occ}$ of Y in $6c_1$ [$AB_5$] | | | | | 0.28 | 0.4 | 1 |
| $N_{occ}$ of Y in $6c_2$ [$AB_2$] | | | | | 0.92 | 1 | 1 |
| $z$ (Y,La) $6c_1$ | | | | | 0.0514 (2) | 0.0510 (1) | 0.0598 (1) |
| $z$ (Y,La) $6c_2$ | | | | | 0.1489 (2) | 0.1482 (1) | 0.1567 (2) |
| $x$ Ni1 $18h$ | | | | | 0.5036 (7) | 0.4997 (5) | 0.5151 (6) |
| $z$ Ni1 $18h$ | | | | | 0.1111 (1) | 0.1096 (1) | 0.1097 (2) |
| $z$ Ni3 $6c_3$ | | | | | 0.2823 (3) | 0.2798 (2) | 0.2700 (2) |
| $z$ Ni4 $6c_4$ | | | | | 0.3847 (3) | 0.3869 (2) | 0.3960 (2) |

* Compound annealed 3 days at 1223 K. Other compounds were annealed 7 days at 1223-1273 K.

### 3.2. Magnetic properties

The thermomagnetic curves $M(T)$ measured under an applied field of 0.1 T are plotted in Figure 5. A distinct behavior is observed for $x < 1$ and $x \geq 1$. To see more clearly the evolution for $x < 1$, the corresponding curves are presented separately with appropriate scales. All the isothermal magnetization curves recorded at 5 K are compared in Figure 6. The magnetic ordering temperatures and magnetic moments are reported in Table 4. The influence of the few impurities observed by XRD on the intrinsic magnetic properties is discussed in the supplementary materials.

The magnetic curves of $La_2Ni_7$ agree with those of previous works showing an antiferromagnetic behavior and a metamagnetic transition towards a wFM state [14, 17, 34]. At 5 K, the critical field $\mu_0 H_c$ obtained from the maximum curve derivative $dM/dH$ is equal to 4.6 T and the ferromagnetic state is reached at 6.5 T, these values are similar to those measured by Parker et *al.* at 4.2 K [12]. The small spontaneous magnetization $M_{spont} = 0.095$ $\mu_B$/f.u. at 5 K (0.048 $\mu_B$ at 300 K) can be attributed to a small quantity of ferromagnetic Ni particles (< 0.1 %), such impurities were not present in the $M_T(\mu_0H)$ curves in refs [12, 17] but observed in ref. [13]. The saturation magnetization extrapolated at high field and corrected from the ferromagnetic Ni impurity is equal to 0.61 $\mu_B$/f.u. or 0.088 $\mu_B$/Ni atom. The thermal evolution of the transition fields, obtained from the derivative of the $M(\mu_0H)$ curves, is presented in Figure 7 and will be discussed later. The $M(T)$ curves have been measured under different applied fields from 0.05 to 7 T (Supplementary, Figure S2a and S2b): a close examination shows that below 1 T the AFM broad peak displays three maxima, as it was observed but not detailed in previous works [12, 14, 15, 17]. Two maxima are still visible between 1 and 3 T and only a broad peak between 3 and 5.5 T. The temperature maxima denoted $T_{N1}$, $T_{N2}$ and $T_{N3}$ decrease *versus* applied field and show thermal hysteresis of 2-4 K between heating and cooling (Supplementary, Figure S2c and S2d). The relative peak intensity also changes *versus* temperature: the intensity of peaks 1 and 2 progressively decreases at the expense of the peak 3 which remains visible up to 5.5 T. The $I_{peak2}/I_{peak3}$ ratio decrease *versus* applied field is shown in inset of Figure S2c.

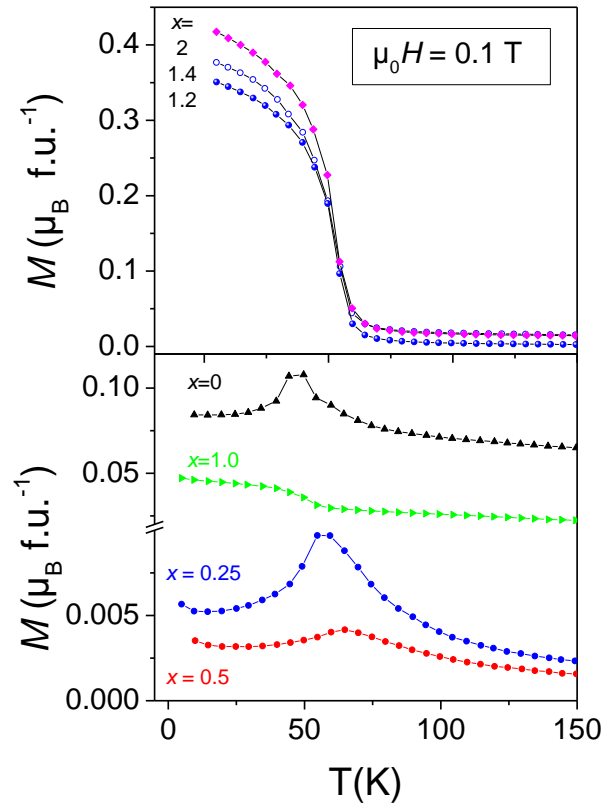

**Figure 5**. Thermomagnetic curves of $La_{2-x}Y_xNi_7$ compounds under an applied field of 0.1 T.

A tentative $\mu_0H = f(T)$ phase diagram was proposed by Tazuke *et al.* [17] using the isothermal $M(\mu_0H)$ curves measured at different temperatures. They have assumed the existence of one AFM phase below transition fields $\mu_0H_{c1}$ ($T \leq 37$ K) and $\mu_0H_{c2}$ (37 K $\leq T <$ 54 K) and two intermediate phases $IM_1$ ($T \leq 37$ K) and $IM_2$ (37 K $< T <$ 66 K) below the third transition field $\mu_0H_{c3}$, above which a FM behavior is found. It can be noticed that the three temperature maxima $T_{Nn}$ (n = 1 to 3) measured upon heating are located at 38 K, 52.2 K and 65 K respectively for an applied field of 0.05 T, and are close to the temperature boundaries observed by Tazuke *et al.* [17].

This reveals a complex magnetic behavior of La$_2$Ni$_7$ which origin has still to be solved. The co-existence of different AFM ordering temperatures can be due for example to a complex helimagnetic structure with propagation vectors along the different crystallographic axis as it has been observed for holmium-based compounds [50-52].

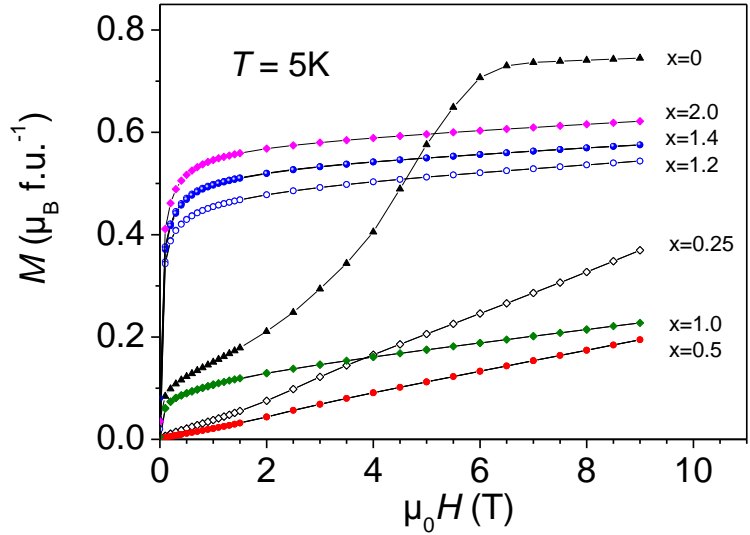

**Figure 6**. Magnetization curves of La$_{2-x}$Y$_x$Ni$_7$ compounds *versus* applied field at 5 K.

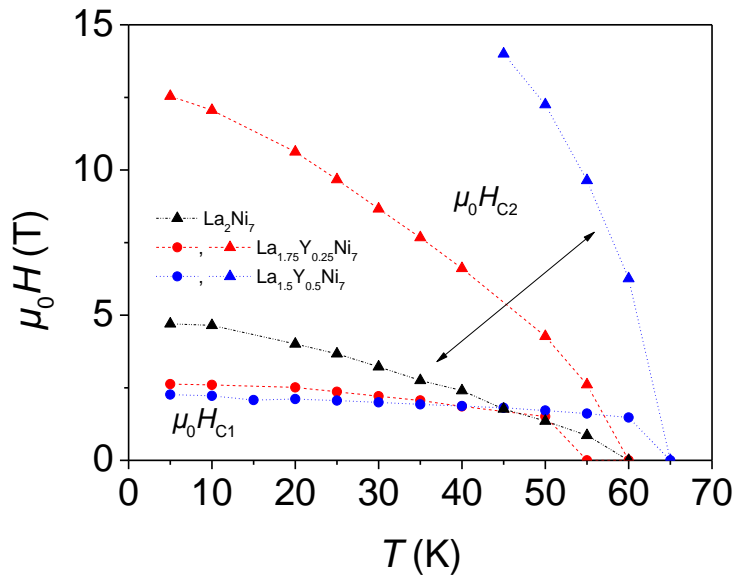

**Figure 7**. Transition fields of La$_{2-x}$Y$_x$Ni$_7$ compounds ($x = 0$, 0.25 and 0.5) *versus* temperature.

For $x = 0.25$ and 0.5, a single AFM peak is observed for an applied field of 0.1 T with maxima at 57 K and 65 K respectively (Figure 5). Contrary to La$_2$Ni$_7$, no splitting of the AFM peak can be observed on the thermomagnetic curves recorded in various applied magnetic fields. As the applied field increases, $T_N$ is shifted to lower temperature and disappears at 14 T for $x = 0.25$, whereas AFM peak with a maximum at 48 K is still observed at 14 T for $x = 0.5$ (Figs. 8a and 9a).

The magnetization isotherms $M_T(\mu_0 H)$ of both compounds display metamagnetic behaviors for $T < T_N$ (Figs. 8b and 9b). All the magnetization curves cross zero, indicating the absence of ferromagnetic impurities. The derivatives of the $M_T(\mu_0 H)$ curves reveal two transition fields $\mu_0 H_{c1}$ and $\mu_0 H_{c2}$ (Supplementary: as example see Figure S3). $\mu_0 H_{c1}$ saturates at about 2.5 T at 5 K for both substituted compounds. For $La_{1.75}Y_{0.25}Ni_7$, $\mu_0 H_{c2}$ increases continuously as $T$ decreases reaching 12.5 T at 5 K. This later metamagnetic transition shows a weak hysteresis below 20 K. For $La_{1.5}Y_{0.5}Ni_7$, the second transition field $\mu_0 H_{c2}$ could only be measured for 45 K $< T <$ 60 K as below 45 K, the transition fields are larger than 14 T. The evolution of the transition fields *versus* temperature for $x = 0$, 0.25 and 0.5 has been plotted in Figure 7. The first transition field $\mu_0 H_{c1}$ around 2 T is barely sensitive to the temperature and the Y content. The second transition field $\mu_0 H_{c2}$ strongly varies *versus* temperature and Y content. $\mu_0 H_{c2}(T)$ increases *versus* temperature, and a scaling factor can be established: it is 2.65 time larger for $x = 0.25$ than for $La_2Ni_7$ and 6 time larger for $x = 0.5$. By extrapolation, $\mu_0 H_{c2}$ should reach about 28 T at 5 K for $x = 0.5$. In addition, for $x = 0.25$, we observe that $T_N = f(\mu_0 H)$ is equivalent to $\mu_0 H_{c2} = f(T)$. The increase of $T_N$ and $\mu_0 H_{c2}$ confirms the stabilization of the AFM state compared to the FM state as the Y content increases. The first transition, between two AFM structures can be related to a spin-flop transition with a partial spin reorientation of the Ni moments upon applying magnetic field, whereas the second transition towards a FM state can be attributed to a spin-flip, with all Ni moments aligned parallel to the applied field. Such behavior has been proposed to describe the $\mu_0 H = f(T)$ phase diagram of metamagnetic compounds [53]. In this study, the existence of two metamagnetic transitions, a first order spin-flop transition with a constant transition field and a second spin-flip transition with a critical field decreasing with the temperature, has been attributed to centrosymmetric system with a weak anisotropy. The similarity of the proposed $\mu_0 H = f(T)$ phase diagram shape (see Figure4 of ref. [53]) with our experimental results strongly suggest that hexagonal $La_{2-x}Y_xNi_7$ compounds ($x < 1$), which crystallize in a centrosymmetric structure, are weak itinerant antiferromagnets with a weak anisotropy.

This metamagnetic behavior can be attributed to itinerant electron metamagnetic transition associated to large spin fluctuations with competition between antiferromagnetic and ferromagnetic states.

The nature of the magnetic transitions can be estimated using the relation between the applied field and magnetization $M$, which is derived from the Landau development of the free energy at equilibrium:

$$\mu_0 H = a_1(T).M + a_2(T)M^3 + a_3(T)M^5 \qquad (1)$$

where $a_1$, $a_2$ and $a_3$ are the Landau coefficients.
The first Landau coefficient $a_1$ corresponds to the inverse of the magnetic susceptibility, and is positive and maximum at $T_C$ or $T_N$. The sign of the second coefficient $a_2$ at the transition temperature indicates the nature of the transition: it is positive or equal to zero for a second order transition and negative for a first order transition at the transition temperature. The sign of the $a_2$ coefficient can be easily determined by plotting the Arrot Belov curves:

$$M^2 = f(H/M) \qquad (2)$$

The Arrott-Belov plots for $x = 0.25$ and 0.5 show below $T_N$, two S-shape behavior corresponding to the two metamagnetic transitions (Figures 8c and 9c and zooms in Figures S4 and S5). They both display negative slopes around $\mu_0 H_{c1}$ and $\mu_0 H_{c2}$ showing the first order character of these transitions. The change from positive to negative slope is clearly observed between 60 and 55 K for $x = 0.25$ and 65 and 60 K for x=0.5, confirming the slight increase of $T_N$ versus Y concentration.

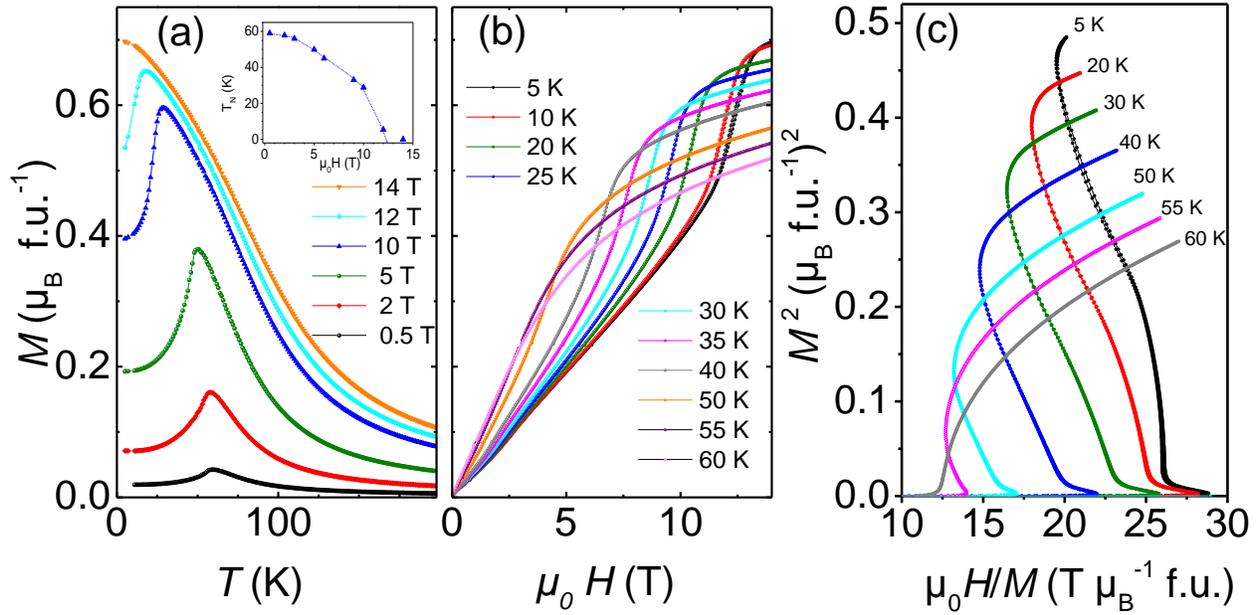

**Figure 8**. Magnetization curves of La$_{1.75}$Y$_{0.25}$Ni$_7$ compound (a) $M_H(T)$ *versus* temperature, (b) $M(\mu_0H)$ *versus* applied field between 5 and 60 K and (c) Arrott-Belov plots. Inset of (a): $T_N$ *versus* applied field.

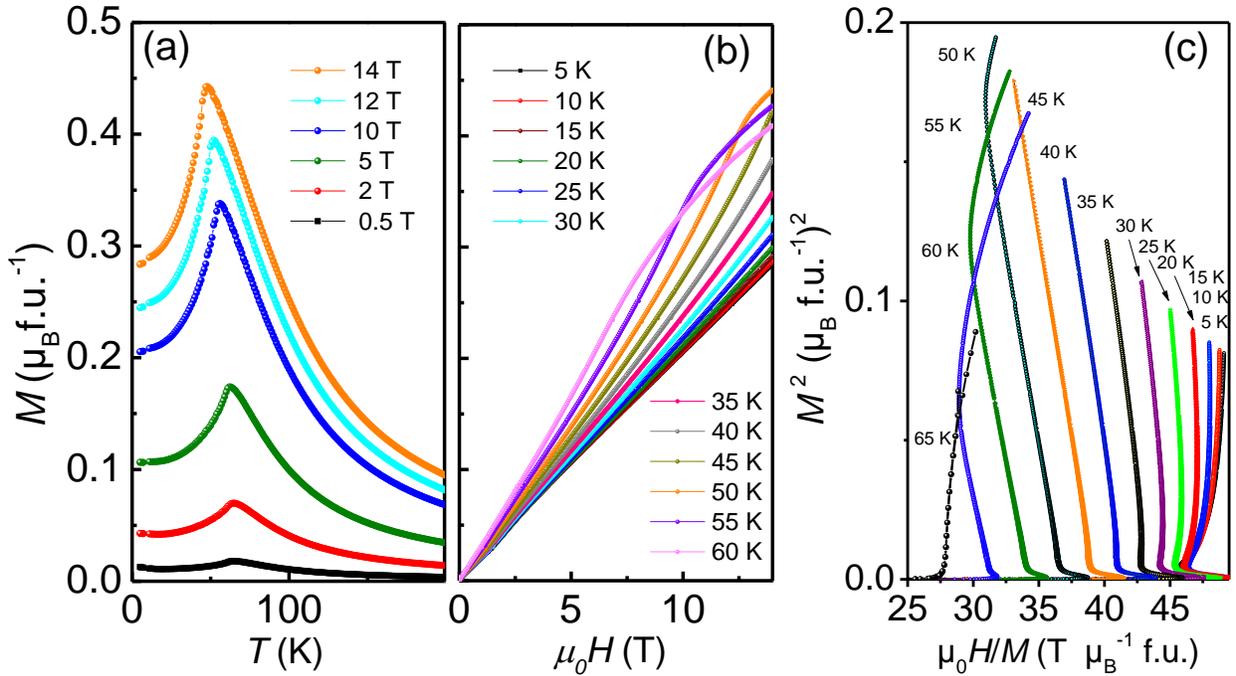

**Figure 9**. Magnetization curves of La$_{1.5}$Y$_{0.5}$Ni$_7$ compound (a) $M_H(T)$ *versus* temperature, (b) $M(\mu_0H)$ *versus* applied field between 5 and 60 K and (c) Arrott-Belov plots.

LaYNi$_7$ displays a very weak ferromagnetic behavior at low magnetic field, and an AFM behavior as the applied field increases (Figure S6). LaYNi$_7$ has a single hexagonal structure, but Y atoms are substituted to La atoms on both *A* sites. This induces some kind of chemical disorder, which can favor a weak ferromagnetic contribution. It is also possible that some traces of rhombohedral phase exist, in too weak percentage to be detected by XRD, but enough to contribute to the magnetization at low field. Nevertheless, as the AFM peak is increasing upon applied field, it confirms that the main magnetic contribution is the AFM one. No metamagnetic transition field can be observed, even for the $M(\mu_0 H)$ curves close to 60 K, indicating that the transition fields are larger than 9 T.

All the $M(T)$ curves of compounds with $x > 1$ present a wFM behavior with $T_C$ around 53 K (Figure 5a). The magnetization isotherms $M_T(\mu_0 H)$ are characteristic of a ferromagnetic behavior (Supplementary material, Figure S7). The incomplete saturation of the $M_{4.2K}(\mu_0 H)$ curve has been already observed for Y$_2$Ni$_7$ and attributed to the itinerant character of the ferromagnetism [22]. The main difference for the three compounds is the difference of extrapolated saturation magnetization ($\mu_S$) which increases with the Y content (Figure 6).

The results obtained for rhombohedral Y$_2$Ni$_7$ agree well with previous works [22, 36, 54]. The smaller $\mu_S$ values observed for $x = 1.2$ and 1.4 can be attributed to the magnetic contribution of the hexagonal phase (33 and 27 wt. %, respectively) which has a smaller magnetization than the rhombohedral one.

In all investigated compounds, the paramagnetic Curie temperatures $\theta_p$ are positive, which is a characteristic of ferromagnets as well as metamagnets, and the effective Ni moments $\mu_{eff}$ vary between 0.8 and 1 $\mu_B$/Ni. For $x = 1.4$, $\mu_{eff} = 0.94$ $\mu_B$/Ni, whereas the saturation moment measured at 5 K is $\mu_S = 0.074$ $\mu_B$/Ni. These values allow to calculate the Rhodes-Wohlfarth ratio (RWR = $q_C/q_S$ ($\mu^2_{eff} = q_C (q_C+2)$; $\mu_S = q_S$)) which characterizes the itinerancy of the ferromagnetic state [22, 55, 56]. For La$_{2-x}$Y$_x$Ni$_7$ compounds ($x = 0, 0.25$), the $\mu_S$ values were estimated by extrapolation of the high magnetic field measurements. The large values of RWR confirm the itinerant magnetic character of all these compounds (Table 4) and the values determined for La$_2$Ni$_7$ and Y$_2$Ni$_7$ agree with previous works [17, 22]. In addition, all these compounds present large ordering temperatures (50 – 60 K) compared to small Ni moments (0.07 – 0.1 $\mu_B$) as expected for weak itinerant magnets.

**Table 4**. Magnetic parameters of the La$_{2-x}$Y$_x$Ni$_7$ compounds.

| $x$ | $T_N$ (K) | $T_C$ (K) | $\mu_S$ at 5 K ($\mu_B$/f.u.) | $\mu_S$ at 5 K ($\mu_B$/Ni) | $\theta_p$ (K) | $\mu_{eff}$ ($\mu_B$/Ni) | $q_C$ ($\mu_B$/Ni) | $q_C/q_S$ |
|---|---|---|---|---|---|---|---|---|
| 0 | 50(2) | | 0.61(1)[b] | 0.088(1) | 67[c] | 0.80[c] | 0.28 | 3.2 |
| 0.25 | 57(2) | | 0.69(5) | 0.099(1) | 60 | 0.82 | 0.30 | 3.1 |
| 0.5 | 65(2) | | | | 53 | 0.80 | 0.28 | |
| 1 | 65(2)[a] | 53(2) | | | 51(2) | 0.88 | 0.33 | |
| 1.2 | | 53(2) | 0.461(7) | 0.066(1) | 52(2) | 1.00 | 0.42 | 6.4 |
| 1.4 | | 54(2) | 0.516(8) | 0.074(1) | 50(2) | 0.94 | 0.39 | 5.3 |
| 2 | | 53(2) | 0.562(9) | 0.080(1) | 50(2) | 0.90 | 0.38 | 4.8 |

[a] For $\mu_0 H > 3$ T; [b] measured at $\mu_0 H > 7$ $\mu_B$; [c] from ref. [17]

The previous experimental results show that the La$_{2-x}$Y$_x$Ni$_7$ compounds which crystallize in a single hexagonal structure with $x < 1$ are weak antiferromagnets with metamagnetic behavior (metamagnets), whereas the one containing a rhombohedral phase with $x > 1$ are weak ferromagnets. LaYNi$_7$, which is just at the boundary, contains at low field a mixture of weak FM and AFM phases. The different magnetic properties of the hexagonal and rhombohedral compounds can be related to their crystal structures and electronic properties. Spin-polarized band structure calculations have been performed for binaries La$_2$Ni$_7$ and Y$_2$Ni$_7$ in both 2$H$ and 3$R$ structures and are detailed in [35]. The calculated non-spin-polarized (NSP) density of state (DOS) of 2$H$-La$_2$Ni$_7$ and 3$R$-Y$_2$Ni$_7$ present a sharp and narrow peak centered at the Fermi level ($E_F$) in agreement with previous studies [11, 23, 57]. These values, which are due to 3$d$ Ni contributions, are large enough, according to Stoner criteria, to stabilize a magnetic state. The wFM structures calculated for both hexagonal and rhombohedral compounds correspond to a distribution of Ni moments which are parallel to the $c$ axis. The Ni moment magnitude are minimum (0.23 to 0.7 $\mu_B$/Ni) for the Ni5 atoms located in the [$AB_2$] units and increase to a maximum (0.3 $\mu_B$/Ni) for the Ni2 atoms at the interface between the [$AB_5$] units. The calculated values of the Ni moments have been explained by a competition between the exchange interactions with the neighboring Ni atoms and the electron charge transfer from the $A$ atom to each Ni atoms.

The AFM structure which has been found for 2$H$-La$_2$Ni$_7$ is described by two ferromagnetic layers of opposite directions separated by a non-magnetic layer at $z = 0.5$ (Figure S8). The Ni5 atoms located in the [$AB_2$] units at $z = 0.5$, which have a very low magnetic moment in the FM structure, lose their ordered moment in the AFM one. The molecular field is canceled on the Ni5 site which is surrounded by 6 Ni1 atoms coupled antiferromagnetically. The metamagnetic AFM-FM transition occurs when the applied magnetic field becomes large enough to induce a Ni moment on the Ni5 site and reverse the orientation of the Ni1 moments from antiparallel toward parallel alignment. The DFT calculations indicate that the FM and AFM structures have the same total energy, therefore the stabilization of the AFM structure should be attributed to more favorable spin fluctuations. The existence of a transition field between two different AFM structures (AFM$_1$ and AFM$_2$) suggests that a spin-flop transition with a change of Ni moment orientation but maintaining an AFM order is induced upon applied field. Unfortunately, neutron powder diffraction (NPD) experiments will not be helpful to solve these two AFM structures as the Ni moments are two small, and their contribution to the NPD pattern negligible. The increase of $T_N$ and of the relative transition fields $\mu_0 H_{c2}$ suggest that the AFM$_2$ phase is stabilized upon Y substitution. As for $x = 0.25$ and $x = 0.5$, the Y atoms are substituted only in the site belonging to the [$AB_2$] unit, they should mainly influence the moments of the Ni5 and Ni1 neighboring atoms, by a charge transfer and a small modification of the DOS at $E_F$. The AFM structure being characterized by the absence of ordered moment on the Ni5 site, its stabilization can originate from a larger charge transfer from the $A$ elements ($A$ = La and Y) towards the Ni5 atom as explained in [35].

Considering the rhombohedral structure, the formation of a similar AFM structure will require at least a doubling of the magnetic cell along the $c$ axis, as an even number of stacking of one [$AB_2$] unit and two [$AB_5$] units is necessary to obtain an equal number of antiparallel FM Ni layers. This would yield a very large anisotropic magnetic cell with $c/a = 14.7$. In addition, the presence of stacking faults in such compounds, which contain a mixture of hexagonal and rhombohedral phases cannot fulfill the periodicity required for the wAFM structure.

All the compounds with $x > 1$, containing a majority of rhombohedral phase, are wFM with the same Curie temperature $T_C = 53$ K, determined by the maximum of the d$M$/d($\mu_0 H$) curves at 0.1 T. It might be surprising that $T_C$ is not affected by the Y for La substitution. However, it should be considered that the magnetic properties arise from the Ni–Ni interactions, which contribute to the sharp and narrow peak in the DOS at $E_F$. In those weak ferromagnets, the magnetic energy is proportional to the energy difference between the ferromagnetic state (SP state) and the paramagnetic state (NSP state). In ref. [35], the calculated energy difference ($\Delta E = E_{SP} - E_{NSP}$) is very close for 3$R$–La$_2$Ni$_7$ (-4.43 meV/Ni atom) and 3$R$–Y$_2$Ni$_7$ (-5 meV/Ni atom), which corresponds to $T_C$ = 51 and 58 K, respectively. This can explain that $T_C$ is not sensitive to the Y for La substitution in the rhombohedral structure for these weak itinerant magnets. The main difference between compounds with $x$ = 1.2, 1.4 and 2 is related to the saturation magnetization, which increases with the Y content. Assuming that the ferromagnetic behavior is related to the rhombohedral phase as observed for Y$_2$Ni$_{7-y}$ compounds [20], the magnetization curves should be a mixture of ferromagnetic and antiferromagnetic phases. However, as the AFM magnetization is much smaller than the FM one, it contributes only to a decrease of the saturation magnetization

Sensitivity of the magnetic order to the chemical composition has been observed upon Al, Co or Cu for Ni substitution in Y$_2$Ni$_7$. The Curie temperature decreases to zero with 1.5 at% of Al and 20 at% of Co [38] or 10 at% of Cu [58]. The collapse of ferromagnetism has been interpreted by a shift of $E_F$ towards the sharp slope of the DOS peak and consequently a fast decrease of the $N(E_F)$ due to these small Ni substitutions.

The influence of applied pressure on the evolution of the magnetic properties of these $A_2$Ni$_7$ compounds could be interesting not only to separate chemical and cell volume influence on their magnetic structures but also to check the possibility to reach a quantum critical point. For example, application of an external pressure on weak itinerant ferromagnet ZrZn$_2$ leads to the observation of a quantum critical point at 2.35 GPa [59].

## 4. CONCLUSION

In the present study, both structural and magnetic properties of the La$_{2-x}$Y$_x$Ni$_7$ system have been investigated. This system is very sensitive to the Y content and two different behaviors are clearly observed for $x \leq 1$ and $x > 1$. For $x \leq 1$, the stable structure is hexagonal, and the Y atoms replace preferentially the La atoms in the [$AB_2$] units of lower coordination number (CN16). This leads to a reduction of the $c/a$ ratio because of geometric constraints between [$AB_2$] and [$AB_5$] units. For compounds with $1 < x < 2$, Y atoms occupy both [$AB_2$] and [$AB_5$] units and the compounds crystallize in a mixture of hexagonal and rhombohedral structures with an increase of the $c/a$ ratio. Y$_2$Ni$_7$ is purely rhombohedral.

The magnetic ground state changes from antiferromagnetic state for compounds with hexagonal structure and $x < 1$ to ferromagnetic state for compounds containing a rhombohedral phase with $x > 1$. La$_2$Ni$_7$ exhibits a complex magnetic phase diagram as its $M(T)$ curves display three AFM peaks, whose positions and relative intensities vary upon applying field. For $x$ = 0.25 and 0.5, at least two itinerant metamagnetic transitions have been observed below $T_N$: a first one between two AFM structures and the second one corresponding to an AFM-FM transition. The critical field $\mu_0H_{c2}$ corresponding to an AFM-FM transition increases with the substituted Y content, indicating a stabilization of the AFM phase. This result also confirms that the AFM structure is stabilized for $A_2$Ni$_7$ compounds crystallizing in the hexagonal structure and with Y substituted only in the [$AB_2$] unit. 2$H$–LaYNi$_7$ compound has mainly an antiferromagnetic behavior but shows some weak ferromagnetic contribution, at low field, which can be related to the substitution of Y on both [$AB_2$] and [$AB_5$] sites.
The rhombohedral compounds with $x > 1$ are all weak ferromagnets with $T_C$ close to 53 K. The Curie temperature remains constant versus Y content indicating very similar magnetic Ni-Ni exchange interaction and only small variations of the saturation magnetization are observed.
These results highlight the correlation of the magnetic properties with the structural changes induced by the Y for La substitution in La$_2$Ni$_7$. Further works will be undertaken to determine the influence of the La substitution by a magnetic rare earth on both structural and magnetic properties of these compounds.

ACKNOWLEDGMENTS

We are thankful to M. Warde for her participation in the synthesis of some compounds presented in this work. We also thank E. Leroy for the EPMA analysis and F. Maccari for the magnetic measurements of $La_{1.5}Y_{0.5}Ni_7$ up to 14 T at TU Darmstadt (Germany). DFT calculations were performed using HPC resources from GENCI-CINES (Grant 2019-96175).

**References**

[1] Buschow KHJ and Van Der Goot AS 1970 The crystal structure of rare-earth nickel compounds of the type $R_2Ni_7$ *J. Less-Common Met.* **22** 419-28

[2] Tazuke Y, Nakabayashi R, Murayama S, Sakakibara T and Goto T 1993 Magnetism of $R_2Ni_7$ and $RNi_3$ (R=Y, La, Ce) *Physica B* **186-188** 596-98

[3] Yartys VA, Vajeeston P, Riabov AB, Ravindran P, Denys RV, Maehlen JP, Delaplane RG and Fjellvag H 2008 Crystal chemistry and metal-hydrogen bonding in anisotropic and interstitial hydrides of intermetallics of rare earth (R) and transition metals (T), $RT_3$ and $R_2T_7$ *Z. Kristall.* **223** 674-89

[4] Srivastava S and Srivastava ON 1999 Hydrogenation behaviour with regard to storage capacity, kinetics, stability and thermodynamic behaviour of hydrogen storage composite alloys, $LaNi_5/La_2Ni_7$, $LaNi_3$ *J. Alloys Compds* **290** 250-56

[5] Charbonnier V, Zhang J, Monnier J, Goubault L, Bernard P, Magen C, Serin V and Latroche M 2015 Structural and Hydrogen Storage Properties of $Y_2Ni_7$ Deuterides Studied by Neutron Powder Diffraction *J. Phys. Chem. C* **119** 12218-25

[6] Gao ZJ and Zhang HM 2016 $(La_{1.66}Mg_{0.34})Ni_7$-based alloys : Structural and Hydrogen Storage Properties *Int. J. Electrochem. Sci.* **11** 1282-92

[7] Gal L, Charbonnier V, Zhang J, Goubault L, Bernard P and Latroche M 2015 Optimization of the La substitution by Mg in the $La_2Ni_7$ hydride-forming system for use as negative electrode in Ni-MH battery *Int. J. Hydr. Energ.* **40** 17017-20

[8] Puga B, Joiret S, Vivier V, Charbonnier V, Guerrouj H, Zhang J, Monnier J, Fariaut-Georges C, Latroche M, Goubault L and Bernard P 2015 Electrochemical Properties and Dissolution Mechanism of $A_2Ni_7$ Hydrides (A=Y, Gd, La-Sm) *ChemElectroChem* **2** 1321-30

[9] Charbonnier V, Monnier J, Zhang J, Paul-Boncour V, Joiret S, Puga B, Goubault L, Bernard P and Latroche M 2016 Relationship between $H_2$ sorption properties and aqueous corrosion mechanisms in $A_2Ni_7$ hydride forming alloys (*A* = Y, Gd or Sm) *J. Power Sources* **326** 146-55

[10] Dhaouadi H, Ajlani H, Zormati H and Abdellaoui M 2018 Elaboration and electrochemical characterization of two hydrogen storage alloy types: $LaNi_{3-x}Mn_xCr_2$ (x=0, 0.1, and 0.3) and $La_2Ni_7$ *Ionics* **24** 2017-27

[11] Werwinski M, Szajek A, Marczynska A, Smardz L, Nowak M and Jurczyk M 2018 Effect of substitution La by Mg on electrochemical and electronic properties in $La_{2-x}Mg_xNi_7$ alloys: a combined experimental and ab initio studies *J. Alloy Compd.* **763** 951-59

[12] Parker FT and Oesterreicher H 1983 Magnetic properties of $La_2Ni_7$ *J. Less-Common Met.* **90** 127-36

[13] Tazuke Y, Nakabayashi R, Murayama S, Sakakibara T and Goto T 1993 Magnetism of $R_2Ni_7$ and $RNi_3$ (R=Y, La, Ce) *Physica B* **186-188** 596-98

[14] Tazuke Y, Abe M and Funahashi S 1997 Magnetic properties of La-Ni system *Physica B* **237-238** 559-60

[15] Fukase M, Tazuke Y, Mitamura H, Goto T and Sato T 1999 Successive metamagnetic transitions in hexagonal $La_2Ni_7$ *J. Phys. Soc. Jpn.* **68** 1460-61

[16] Fukase M, Tazuke Y, Mitamura H, Goto T and Sato T 2000 Itinerant electron magnetism of hexagonal $La_2Ni_7$ *Mater. Trans.* **41** 1046-51

[17] Tazuke Y, Suzuki H and Tanikawa H 2004 Metamagnetic transitions in hexagonal $La_2Ni_7$ *Physica B* **346** 122-26

[18] Buschow KHJ 1984 Magnetic-properties of $Y_2Ni_7$ and its hydride *J. Less-Common Met.* **97** 185-90
[19] Shimizu M, Inoue J and Nagasawa S 1984 Electronic-Structure and Magnetic-Properties of Y-Ni Intermetallic Compounds *J. Phys. F Met. Phys.* **14** 2673-87
[20] Tazuke Y, Nakabayashi R, Hashimoto T, Miyadai T and Murayama S 1992 Magnetism of Ni-based alloys: weak ferromagnetic and paramagnetic alloys *J. Magn. Magn. Mater.* **104-107** 725-26
[21] Zhou GF, Deboer FR and Buschow KHJ 1992 Magnetic coupling in rare-earth nickel compounds of the type $R_2NI_7$ *J. Alloy Compd.* **187** 299-303
[22] Bhattacharyya A, Jain D, Ganesan V, Giri S and Majumdar S 2011 Investigation of weak itinerant ferromagnetism and critical behavior of $Y_2Ni_7$ *Phys. Rev. B* **84** 184414
[23] Singh DJ 2015 Electronic structure and weak itinerant magnetism in metallic $Y_2Ni_7$ *Phys. Rev. B* **92** 174403
[24] Lonzarich GG 1984 Band structure and magnetic fluctuations in ferromagnetic or nearly ferromagnetic metals *J. Magn. Magn. Mater.* **45** 43-53
[25] Suzuki K and Masuda Y 1985 Volume magnetostriction in itinerant electron ferromagnetic $Ni_3Al$ system *J. Phys. Soc. Jpn.* **54** 326-33
[26] Aguayo A, Mazin II and Singh DJ 2004 Why $Ni_3Al$ is an itinerant ferromagnet but $Ni_3Ga$ is not *Phys. Rev. Lett.* **92** 147201
[27] Naganuma H, Endo Y, Nakatani R, Kawamura Y and Yamamoto M 2004 Magnetic properties of weak itinerant ferromagnetic ζ-$Fe_2N$ film *Sci. Technol. Adv. Mater.* **5** 83-87
[28] Chen GM, Lin MX and Ling JW 1994 Spin fluctuation effect in the ordered $Fe_2N$ alloy *J. Appl. Phys.* **75** 6293-95
[29] Lin S, Lv HY, Lin JC, Huang YA, Zhang L, Song WH, Tong P, Lu WJ and Sun YP 2018 Critical behavior in the itinerant ferromagnet $AsNCr_3$ with tetragonal-antiperovskite structure *Phys. Rev. B* **98** 014412
[30] Yoshinaga S, Mitsui Y, Umetsu RY, Uwatoko Y and Koyama K 2017 Magnetic properties of weak itinerant electron ferromagnet CrAlGe *J. Phys.: Conf. Ser.* **868** 012018
[31] Goh WF and Pickett WE 2017 Competing magnetic instabilities in the weak itinerant antiferromagnetic TiAu *Phys. Rev. B* **95** 205124
[32] Solontsov AZ and Silin VP 2004 Weak itinerant-electron antiferromagnetism of uranium nitride *Phys. Metals Metallogr.* **97** 571-81
[33] Guo GY, Wang YK and Hsu LS 2002 First-principles and experimental studies of the electronic structures and magnetism in $Ni_3Al$, $Ni_3Ga$ and $Ni_3In$ *J. Magn. Magn. Mater.* **239** 91-93
[34] Buschow KHJ 1983 Magnetic properties of $La_2Ni_7$ and its hydride *J. Magn. Magn. Mater.* **40** 224-26
[35] Crivello J-C and Paul-Boncour V 2020 Relation between the weak itinerant magnetism in $A_2Ni_7$ compounds (A = Y, La) and their stacked crystal structures *J. Phys.: Cond. Matter* **32** 145802
[36] Buschow KHJ 1984 Magnetic-properties of $Y_2Ni_7$ and its hydride *J. Less-Common Met.* **97** 185-90
[37] Ballou R, Gorges B, Molho P and Rouault P 1990 Thermal spontaneous magnetization in $Y_2Ni_7$ - a misinterpretation *J. Magn. Magn. Mater.* **84** L1-L4
[38] Dubenko IS, Levitin RZ, Markosyan AS, Petropavlovsky AB and Snegirev VV 1990 Rapid Suppression of Ferromagnetism in $Y_2Ni_7$ and $YNi_3$ Intermetallic Compounds for Small Substitutions of Ni by Al and Co *J. Magn. Magn. Mater.* **90-91** 715-18
[39] Levitin RZ, Markosyan AS, Petropavlovskii AB and Snegirev VV 1990 Existence of temperature-induced ferromagnetism in $Y_2Ni_7$ and influence of minor replacements of nickel by aluminum on the magnetic-properties of this compound *Jetp Lett.* **51** 56-59
[40] Rodriguez-Carvajal J 1993 Recent Advances in Magnetic Structure Determination by Neutron Powder Diffraction *Physica B* **192** 55-69
[41] Crivello JC, Zhang J and Latroche M 2011 Structural Stability of AB(y) Phases in the (La,Mg)-Ni System Obtained by Density Functional Theory Calculations *J. Phys. Chem. C* **115** 25470-78
[42] Kresse G and Hafner J 1993 Ab initio molecular dynamics for open-shell transition metals *Phys. Rev. B* **48** 13115

[43] Kresse G and Joubert D 1999 From ultrasoft pseudopotentials to the projector augmented-wave method *Phys. Rev. B* **59** 1758

[44] Perdew JP, Burke K and Ernzerhof M 1997 Generalized Gradient Approximation Made Simple *Phys. Rev. Lett.* **78** 1396-96

[45] Serin V, Zhang J, Magén C, Serra R, Hÿtch MJ, Lemort L, Latroche M, Ibarra MR, Knosp B and Bernard P 2013 Identification of the atomic scale structure of the $La_{0.65}Nd_{0.15}Mg_{0.20}Ni_{3.5}$ alloy synthesized by spark plasma sintering *Intermetallics* **32** 103-08

[46] Zhang J, Latroche M, Magen C, Serin V, Hÿtch MJ, Knosp B and Bernard P 2014 Investigation of the Phase Occurrence, H Sorption Properties, and Electrochemical Behavior in the Composition Ranges $La_{0.75-0.80}Mg_{0.30-0.38}Ni_{3.67}$ *J. Phys. Chem. C* **118** 27808-14

[47] Casas-Cabanas M, Reynaud M, Rikarte J, Horbach P and Rodríguez-Carvajal J 2016 FAULTS: A program for refinement of structures with extended defects *J. Appl. Cryst.* **49** 2259-69

[48] Ferey A, Cuevas F, Latroche M, Knosp B and Bernard P 2009 Elaboration and characterization of magnesium-substituted $La_5Ni_{19}$ hydride forming alloys as active materials for negative electrode in Ni-MH battery *Electrochimica Acta* **54** 1710-14

[49] Lemort L, Latroche M, Knosp B and Bernard P 2011 Elaboration and Characterization of New Pseudo-Binary Hydride-Forming Phases $Pr_{1.5}Mg_{0.5}Ni_7$ and $Pr_{3.75}Mg_{1.25}Ni_{19}$: A Comparison to the Binary $Pr_2Ni_7$ and $Pr_5Ni_{19}$ Ones *J. Phys. Chem. C* **115** 19437-44

[50] Steinitz MO, Kahrizi M and Tindall DA 1987 Splitting of the Néel transition in holmium in a magnetic-field *Phys. Rev. B* **36** 783-84

[51] Willis F, Ali N, Steinitz MO, Kahrizi M and Tindall DA 1990 Observation of transitions to spin-slip structures and splitting of the Néel temperature of holmium in magnetic fields *J. Appl. Phys.* **67** 5277-79

[52] Tindall DA, Steinitz MO and Holden TM 1993 Studies of the magnetic phase-diagram of holmium using neutron-diffraction *J. Appl. Phys.* **73** 6543-45

[53] Sokolov DA, Kikugawa N, Helm T, Borrmann H, Burkhardt U, Cubitt R, White JS, Ressouche E, Bleuel M, Kummer K, Mackenzie AP and Rößler UK 2019 Metamagnetic texture in a polar antiferromagnet *Nature Physics* **15** 671-77

[54] Nishihara Y and Ogawa S 1991 Itinerant Electron Ferromagnetism in $Y_2Ni_7$ and Mossbauer-Effect of $Fe^{57}$ Doped in $Y_2Ni_7$ *J. Phys. Soc. Jpn.* **60** 300-03

[55] Nakabayashi R, Tazuke Y and Murayama S 1992 Itinerant Electron Weak Ferromagnetism in $Y_2Ni_7$ and $YNi_3$ *J. Phys. Soc. Jpn.* **61** 774-77

[56] Rhodes P and Wohlfarth EP 1963 Effective Curie - Weiss Constant of Ferromagnetic Metals and Alloys *Proc. R. Soc. London Ser. A* **273** 247-58

[57] Werwiński M, Szajek A, Marczyńska A, Smardz L, Nowak M and Jurczyk M 2019 Effect of Gd and Co content on electrochemical and electronic properties of $La_{1.5}Mg_{0.5}Ni_7$ alloys: A combined experimental and first-principles study *J. Alloy Compd.* **773** 131-39

[58] Levitin RZ, Markosyan AS, Petropavlovskii AB and Snegirev VV 1997 On the magnetic properties of small substitutions of iron and copper for nickel in $YNi_3$ and $Y_2Ni_7$ intermetallics *Phys. Solid State* **39** 1633-35

[59] Kabeya N, Maekawa H, Deguchi K, Kimura N, Aoki H and Sato NK 2013 Phase diagram of the itinerant-electron ferromagnet $ZrZn_2$ *Phys. Status Solidi B* **250** 654-56

**Supplementary materials**

The four XRD refined patterns presented in figure S1 are characteristic of the different structures observed in the $La_{2-x}Y_xNi_7$ system.

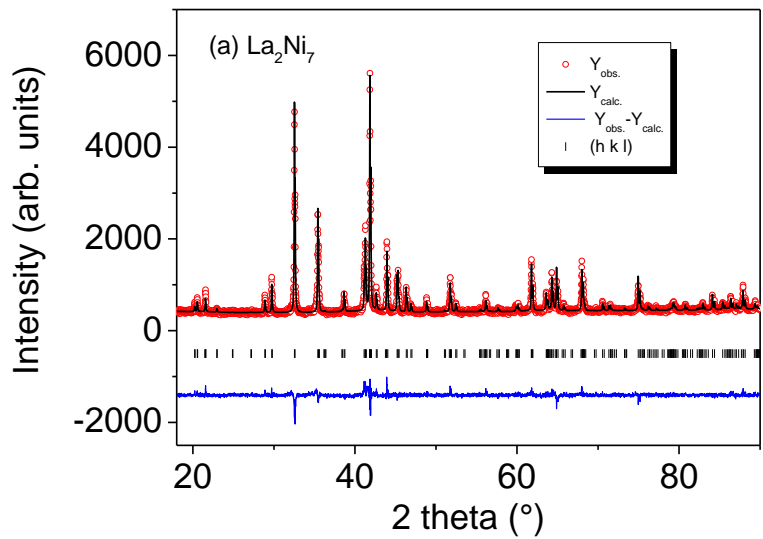

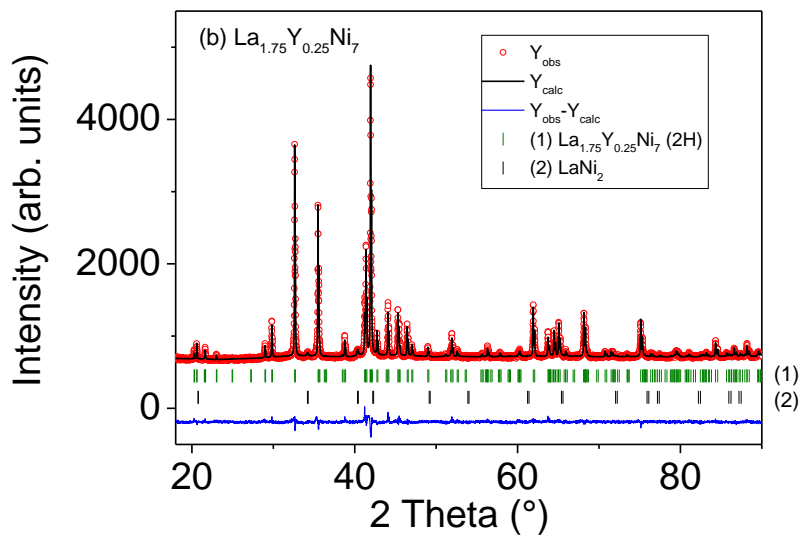

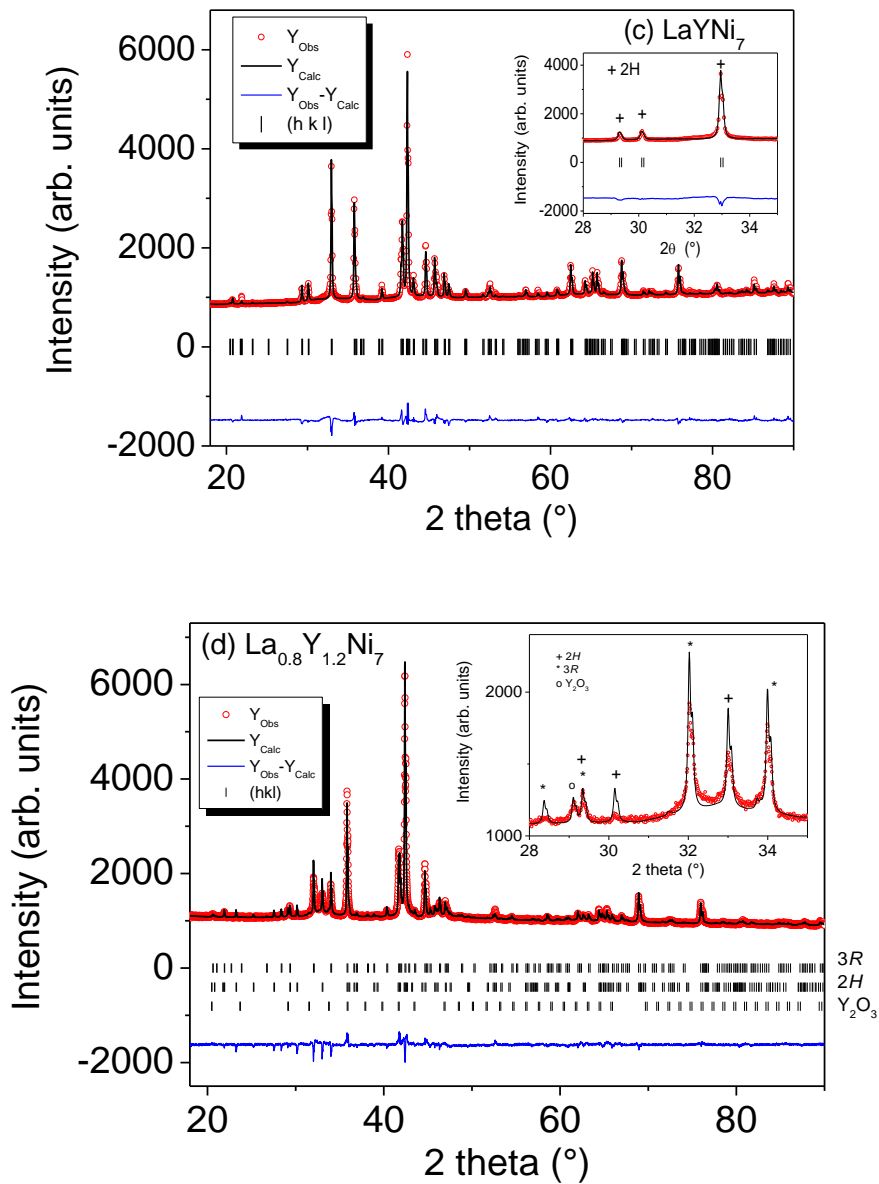

**Figure S1**: Refined XRD patterns of a) $La_2Ni_7$ ($\chi^2 = 1.9$ %, $R_{Bragg} = 10.2$ %), b) $La_{1.75}Y_{0.25}Ni_7$ ($\chi^2 = 2.7$ %, $R_{Bragg} = 9.9$ %), c) $LaYNi_7$ ($\chi^2 = 10$ %, $R_{Bragg} = 9.0$ %) and d) $La_{0.8}Y_{1.2}Ni_7$ ($\chi^2 = 5.2$ %, $R_{Bragg}$ ($3R$) =13.6 %, $R_{Bragg}$ ($2H$) =15.9 % ). Inset of c) and d) zoom showing the 2θ range between 28 and 35 °.

The thermomagnetization curves of $La_2Ni_7$ at different fields are presented in Figure S2, as well as the Néel temperature obtained from the curve maxima.

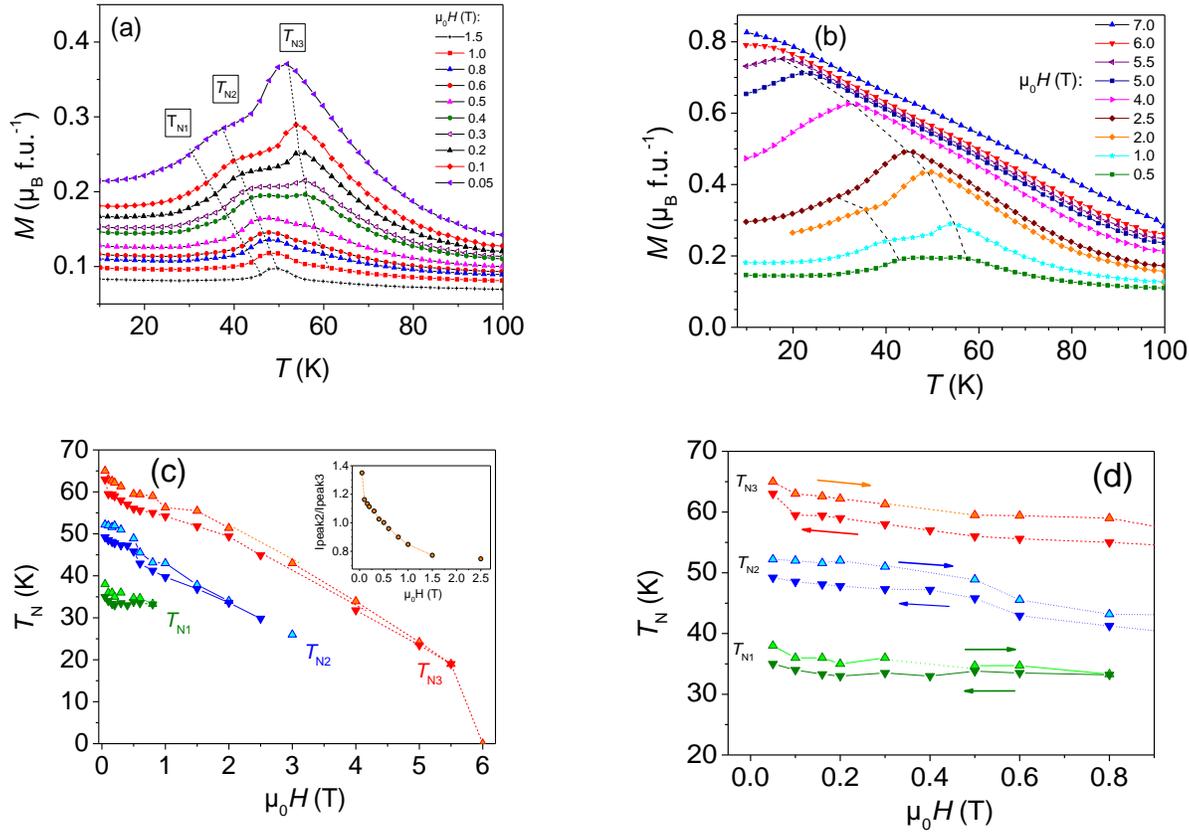

**Figure S2**: (a) and (b) $M(T)$ curves of $La_2Ni_7$ measured upon decreasing temperature versus applied field (c) Néel temperatures $T_{Ni}$ variation upon decreasing and increasing fields and (d) zoom of (c). Inset of (c) ratio of the maxima of peak at $T_{N2}$ versus peak at $T_{N3}$.

The presence of two transition fields are evidenced on the first derivative of the M ($\mu_0H$) curve of La$_{1.75}$Y$_{0.25}$Ni$_7$ at 40 K on Figure S3.

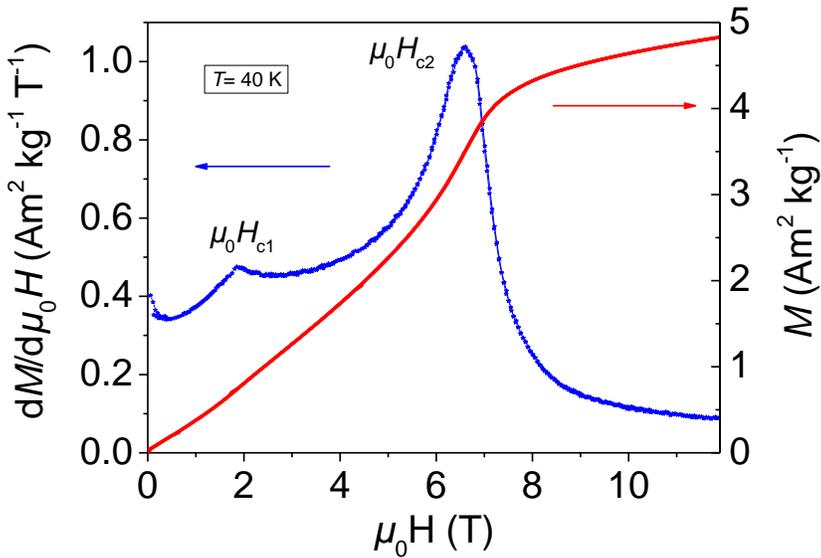

**Figure S3**: Magnetization and first derivative of La$_{1.75}$Y$_{0.25}$Ni$_7$ at 40 K.

The first order character of the spin-flop transition between two AFM structures is evidenced on the Arrot Belov curves of La$_{1.75}$Y$_{0.25}$Ni$_7$ and La$_{1.5}$Y$_{0.5}$Ni$_7$ in Figs. S4 and S5. For both compounds negative slopes have been observed below $T_N$.

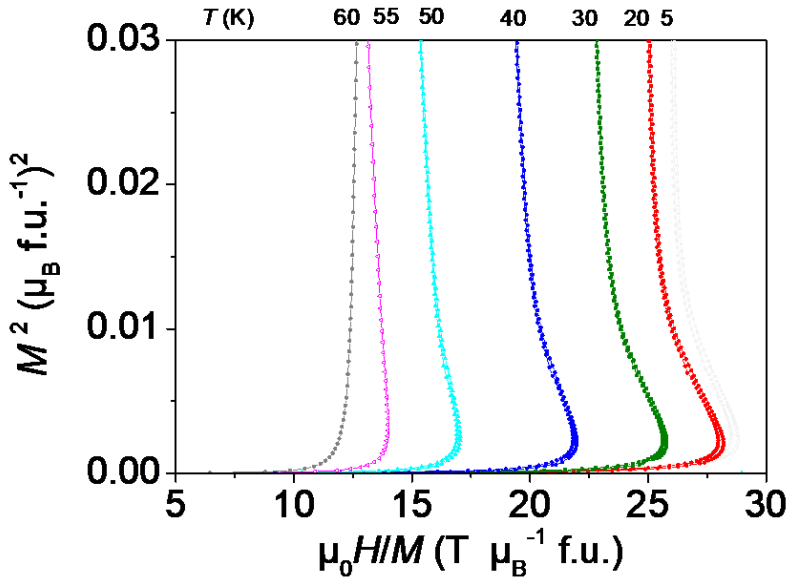

**Figure S4**: Zoom Arrott-Belov plots derived from the La$_{1.75}$Y$_{0.25}$Ni$_7$ magnetization curves.

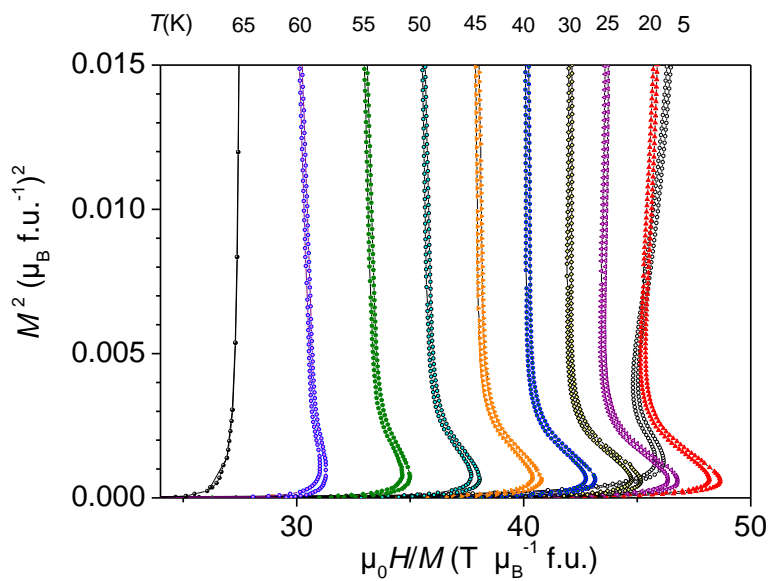

**Figure S5**: Zoom of Arrott-Belov plot derived from the $La_{1.5}Y_{0.5}Ni_7$ magnetization curves.

The magnetizations curves at different temperatures of $LaYNi_7$ and $La_{0.8}Y_{1.2}Ni_7$ are reported in Figs. S6 and S7

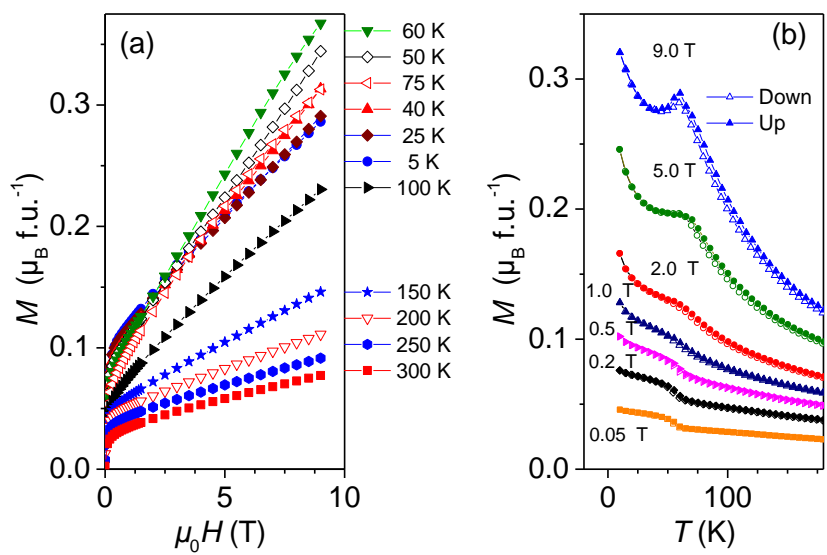

**Figure S6**: Magnetization curves of $LaYNi_7$ versus applied field (a) and temperature (b).

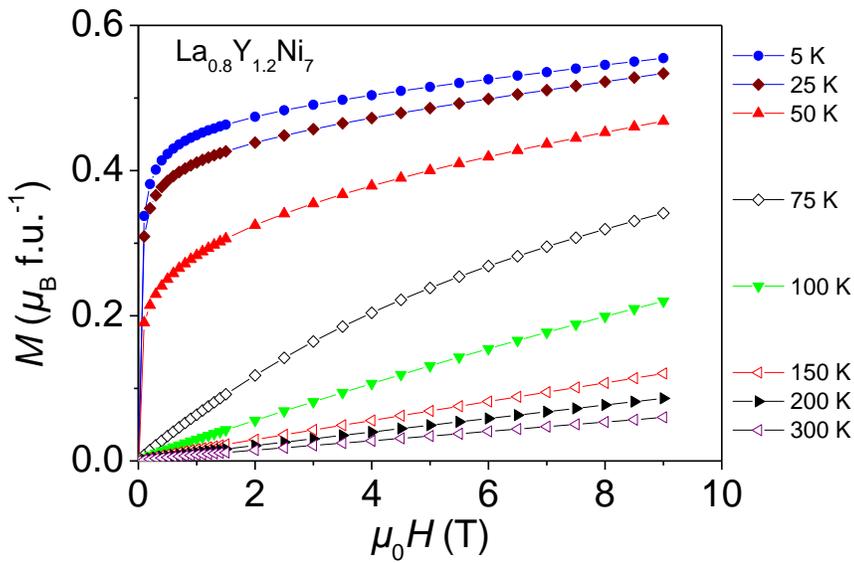

**Figure S7:** Magnetization curves of La$_{0.8}$Y$_{1.2}$Ni$_7$ versus applied field.

Influence of magnetic impurities observed by XRD on the total magnetization.

Note that the presence of 1.4 % of LaNi$_2$ in the XRD pattern of La$_{1.75}$Y$_{0.25}$Ni$_7$, cannot explain the first critical field which is observed not only for La$_{1.75}$Y$_{0.25}$Ni$_7$ but also for La$_{0.5}$Y$_{0.5}$Ni$_7$, as LaNi$_2$ is a Pauli paramagnet with a very weak magnetic susceptibility ($\chi = 1.5 \times 10^{-4}$ emu/mole) (Y. Tazuke, M. Abe, and S. Funahashi, Physica B **237-238**, 559 (1997).

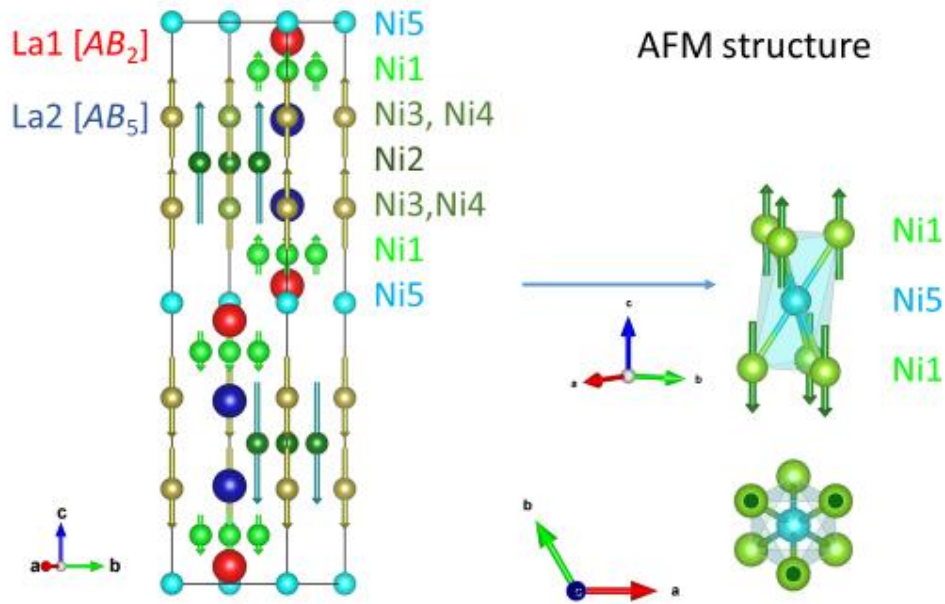

**Figure S8**. Schematic representation of the AFM structure of 2$H$-La$_2$Ni$_7$ [35]. The particular magnetic environment of the Ni5 atoms is detailed along different crystallographic projections.